\documentstyle[psfig]{article}

\begin{document}


\newpage

\noindent
{\LARGE {\bf Nuclei cross sections in Extensive Air Showers}}
\vspace{.5cm}

\noindent
{\Large Tadeusz Wibig and Dorota Sobczy\'{n}ska}
\vspace{.3cm}

{
\large
\noindent
{ \it Experimental Physics Dept., University of \L odz, \\
Pomorska 149/153, 90-236 \L odz, Poland}
\vspace{.5cm}



\noindent
{\bf Abstract}
{Cross sections for proton inelastic collision with different nuclei
are described within the Glauber approximation.
The significant difference between approximate ``Glauber'' formula
and exact calculations with geometrical scaling assumption is shown
for very high energy cross section calculations.
According to obtained results values of proton--proton cross section reported
by the Akeno and Fly's Eye experimental groups are about 10\%
overestimated.
}

\vspace{.2cm}


The rise of the proton-proton cross section (total, inelastic) as the
interaction energy increases is an important feature of the strong
interaction picture. The growth itself is established quite well both
from theoretical and experimental point of view. However the question
how fast do cross sections rise is discussed permanently. A definite answer
is still lacking. Theoretical predictions agree well with one another and
with accelerator data in the region where data exists ($\sqrt{s} \sim 20
\div 2000$ GeV) but they differ above. Before the LHC shifts the
direct measurements limit to 10 $\sim$ 14 TeV the only existing information
can be derived from the cosmic ray extensive air shower (EAS) data.

The important difference between the collider and EAS proton--proton
cross section measurements is that in fact
the proton--air interactions are involved in the EAS development thus
the value which is real measured is the cross section for the interactions
with air nuclei. The value of proton--proton cross section is obtained
from it using a theory for nuclei interactions.
In most
of recent papers concerning this subject it can be found just a few sentences
like ``calculations have been made in the standard Glauber formalism''
or something very similar \cite{akeno,FE}.

The original Glauber paper \cite{glauber} has been published about forty
years ago.
Rather complicated equations for scattering cross sections can
be simplified significantly applying some additional assumptions which
validity is limited.
We would like to compare results of calculations with and without
these simplifications.
The exact Glauber formalism will be used to evaluate
the proton--proton cross section values from the Akeno and Fly's Eye data.

For the scattering of particle on the close many particle system (nucleus),
if each interaction can be treated as a two particle one, the
overall phase shift for incoming wave is a sum of all two-particle
phase shifts.
\begin{equation}
\chi_A(b,\: \{{\bf d}\})~=~\sum_{j=1}^A \: \chi _j ({\bf b}\: -\: {\bf d}_j)
\label{chiskla}
\end{equation}
\noindent
where $\{{\bf d}\}$ is a set of nucleon positions in the nucleus
(${\bf d}_j$ is a position of the $j$th nucleon in the plane
perpendicular to the interaction axis).
The equation (\ref{chiskla}) is the essence of Ref.\cite{glauber}
and in fact it defines the Glauber approximation.
The scattering amplitude is thus given by
\begin{equation}
S(t)~=~{i \over {2 \pi }} \int {\rm e}^{i{\bf t b}} d^2{\bf b}
\int | \psi(\{{\bf d}\} ) | ^ 2 \: \left\{ \: 1 \: - \: {\rm e}
^{i \chi_A(b,\: \{{\bf d}\})}  \right\}
\prod _{j=1}^A d^2 {\bf d}_j~~,
\label{st1}
\end{equation}
\noindent
where $\psi$ describes the wave function of the nucleus with nucleons
distributed according to $\{{\bf d}\}$.
If one neglect position correlations of the nucleons and denotes by
$\varrho _j$ each single nucleon density and
%
if all interactions can be described by the same phase shift
function $\chi$ then
\begin{eqnarray}
S(t)=
{{i} \over {2 \pi }} \int {\rm e}^{i{\bf t b}} d^2{\bf b}
\left\{ 1  -   \int  \prod _{j=1}^A \varrho_j ({\bf d}_j)
{\rm e}^{i \chi({\bf b} - {\bf d}_j)} d^2 {\bf d}_j \right\} ~.
\label{st2}
\end{eqnarray}

On the other hand, the scattering process
can be treated as a one collision process
with its own nuclear phase shift $\chi_{\rm opt}(b)$
\begin{equation}
S(t)~=~{{i} \over {2 \pi }} \int {\rm e}^{i{\bf t b}}
\left\{ 1 \: - \: {\rm e}^{i \chi_{\rm opt}(b)} \right\} d^2 {\bf b}~~.
\label{chiop}
\end{equation}
The comparison with Eq.(\ref{st1}) gives
\begin{eqnarray}
 {\rm e}^{i \chi_{\rm opt}(b)}~=~
\int | \psi(\{{\bf d}\})|^2\:
{\rm e}^{i \sum_{j=1}^A \: \chi _j ({\bf b}\: -\: {\bf d}_j)}
\prod _{j=1}^A d^2 {\bf d}_j
~=~\left\langle {\rm e}^{i \chi(b,\: \{{\bf d}\})} \right\rangle~~,
\label{chiopt}
\end{eqnarray}
\noindent
where the $\langle \ \rangle$ means the averaging over the all possible
configuration of nucleons $\{ {\bf d} \} $.
To go further with the calculations of $\chi_{\rm opt}$ the commonly used
assumption has to be made. If we assume that the number of scattering centers
($A$) is large (with the transparency of the nucleus as a whole constant)
then
\begin{equation}
\chi_{\rm opt}(b)
~=~i\: \int d^2 {\bf d} \rho_A({\bf d})\:
\left[ 1 - {\rm e}^{i \chi({\bf b} - {\bf d})} \right]~.
\label{exact}
\end{equation}
\noindent
where $\rho_A$ is the
distribution of scattering centers (nucleons) positions
in the nucleus ($\sum \varrho_j$).
When the individual nucleon opacity $| 1-{\rm e}^{i \chi(b)} |$ is a very
sharply peaked compared with $\rho_A$ then with the help of the
optical theorem the simple formula for the scattering amplitude can be found.
\noindent
The proton nucleus inelastic cross section is thus
\begin{equation}
\sigma_{pA}^{\rm inel}~=~
\int d^2 {\bf b}
\left[ 1 - {\rm e}^{- \sigma_{pp}^{\rm tot} \rho_A(b)} \right]
~=~
\int d^2 {\bf b}
\left\{ 1 -
\left[ 1-
\sigma_{pp}^{\rm tot} {\rho_A \over A} \right] ^A \right\}
\label{ginel}
\end{equation}
\noindent
This result is often but not quite correctly called
``the Glauber approximation''.

The important point of this paper is to show how the point-nucleon
approximation changes the results.
Formulas given in Eq.
(\ref{ginel}) are in fact in
agreement with the factorization hypothesis for individual nucleon--nucleon
$\chi$ function 
$\chi(s,\:b) ~=~ i\: \omega(b)\:f(s)$. 
According to it nucleons are getting
blacker
as the interaction energy increases.
However, for some time (see, e.g., Ref.\cite{amaldi}) it is known that
this is not the case. Experimental data
strongly favour the geometrical scaling hypothesis
$\chi(s,\:b) ~=~ i\: \omega \left(b\: / b_0(s)\right)$
 (see, e.g., Ref. \cite{buddd})
which treats nucleons
as getting bigger. Nucleus profiles obtained using exact Glauber formula
(Eq.(\ref{exact})] differ from the
$A \sigma_{pp}^{\rm inel} {\rho_A \over A}$ suggested by
Eq.(\ref{ginel}). The difference can be seen clearly
in Fig.~1.


The significant change of the nucleus size have to influence the value
of inelastic cross section. Fig.~2 shows the
change of inelastic cross section of proton--nucleus with the
interaction energy calculated using geometrically scaled and
factorized nucleus profiles.
As it can be seen the difference at very high energies is remarkable.


The results presented above indicate the importance of
reexamination of the proton--proton cross section estimation based
on proton--air data measured in EAS experiment.

The conversion from proton--air to proton--proton cross section
presented in Fig.~3 is obtained using the exact Glauber
formalism.
The original Fly's Eye estimation of proton--proton total cross section
given in Ref.\cite{FE} is 120 mb while according to results given in
Fig.~3 it is equal to 110 mb. The same procedure should be applied
to the Akeno data.


In Fig.~4 the calculated proton--air cross section energy
dependence is given. The solid line represents result
obtained using exact Glauber formula [Eq.(\ref{chiopt})] and proton--proton
phase shift $\chi$ function described in Ref.\cite{xsecprep}.
The outcome of the
``simplified Glauber'' approach [Eq.(\ref{ginel})] is given for a comparison
by the long dashed line.



Concluding, we have shown that
the geometrical scaling hypothesis with the exact Glauber formalism
gives the value of proton--proton total cross section
at about 30 TeV about 10\% smaller than that reported in original
Fly's Eye and Akeno papers. Results presented in this work
have been obtained using the following cross section energy dependence
\begin{equation}
{\sigma_{\rm inel}(s)} = 32.4~-~1.2 \ln (s)~+~0.21 \ln ^2 (s)~~,
\end{equation}
\noindent
which fits quite well accelerator measurements as well as the EAS
points diminish by $\sim$ 10\%.

The rise of the $\sigma_{A-air}$ predicted by
the geometrical scaling hypothesis with the exact Glauber formalism
is significantly faster than the one which can be obtained using
simplified formulas (Fig.~2).
This can change the physical conclusions based on
Monte Carlo simulations of the EAS development at very high energies.


\vspace{.3cm}

\centerline{
\hspace{.8cm}
\psfig{file=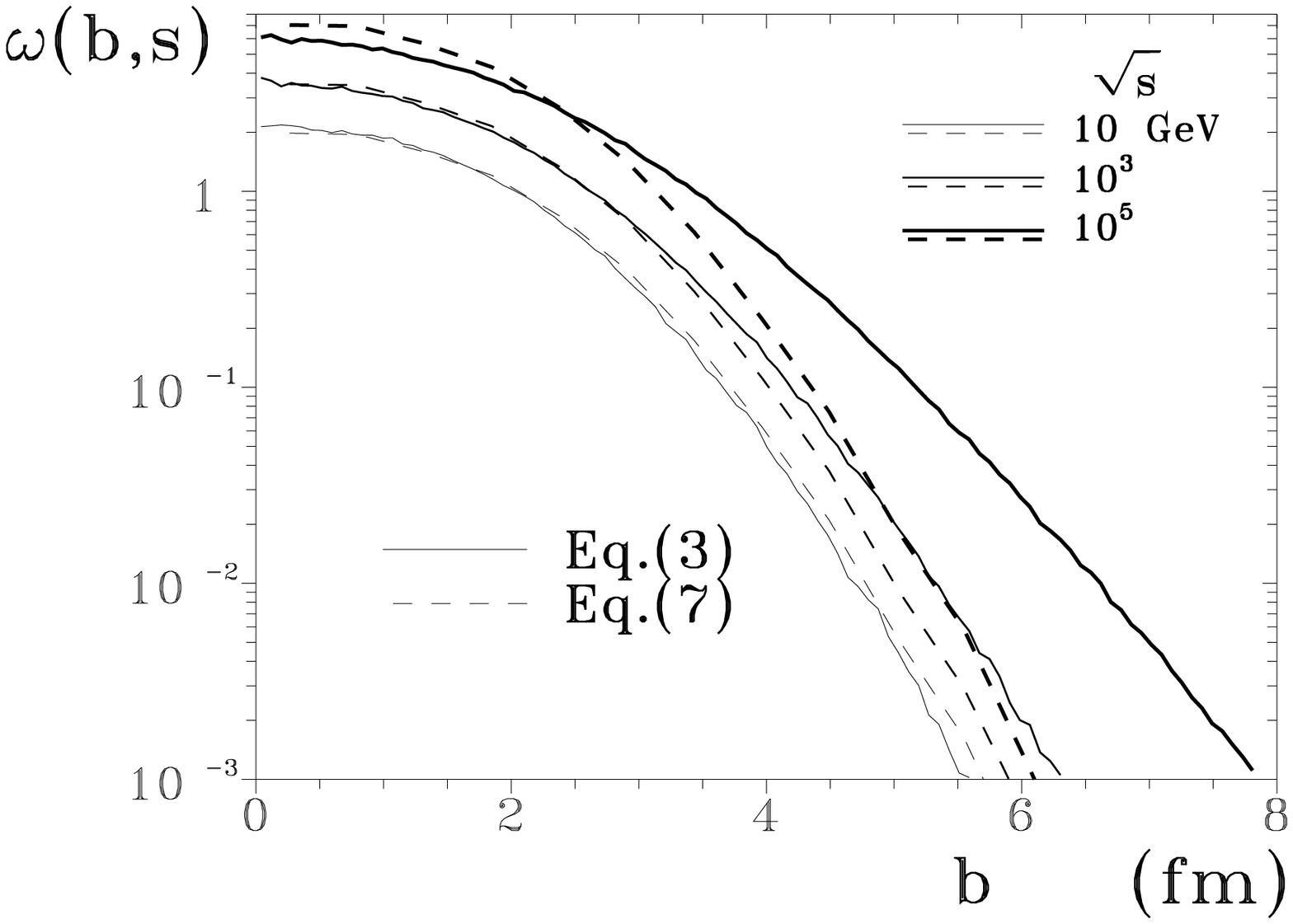,width=8.5cm}
\hspace{-1.cm}
%
\psfig{file=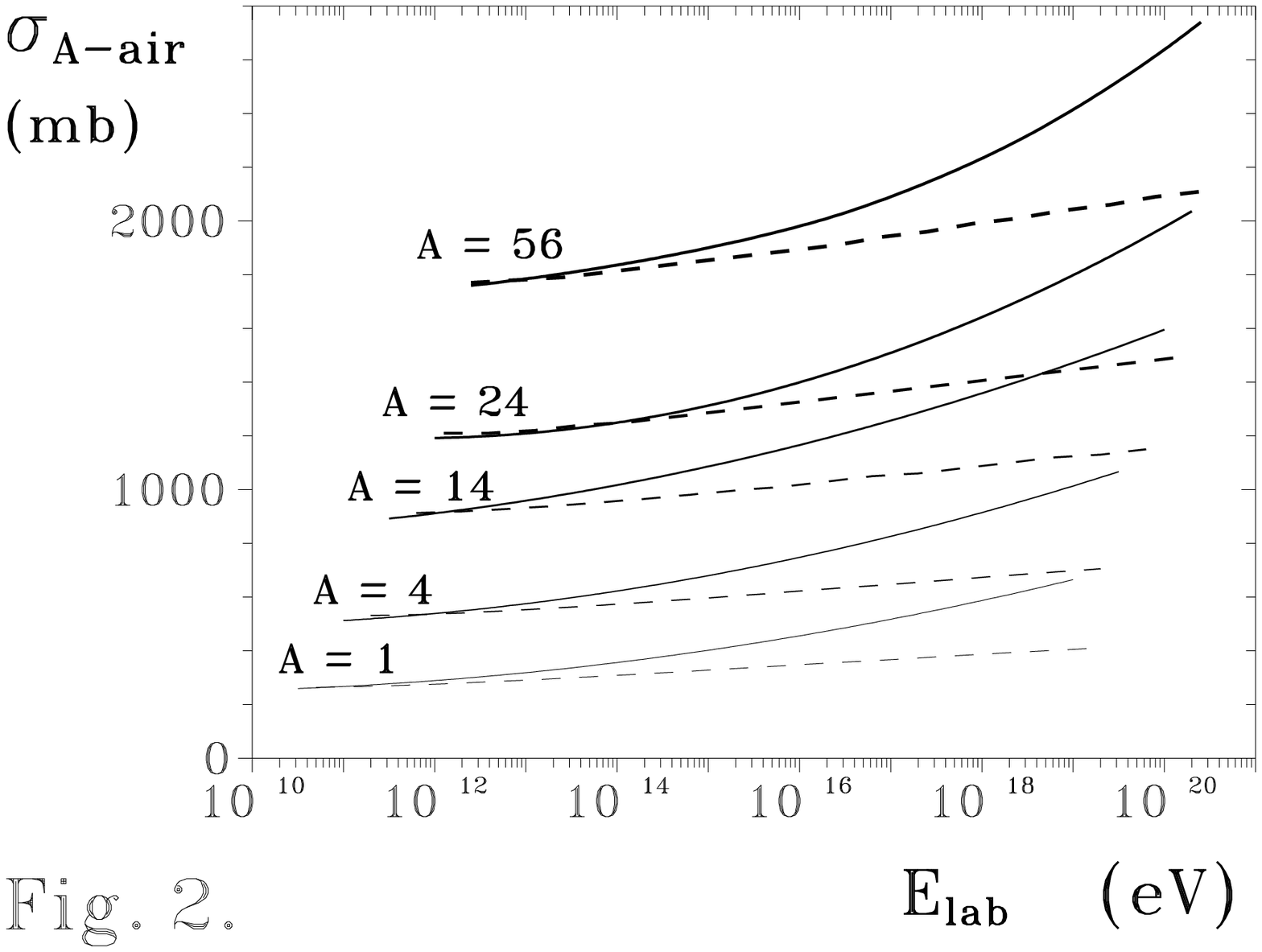,width=8.5cm}}
\vspace{.5cm}

\centerline{
\hspace{.8cm}
\psfig{file=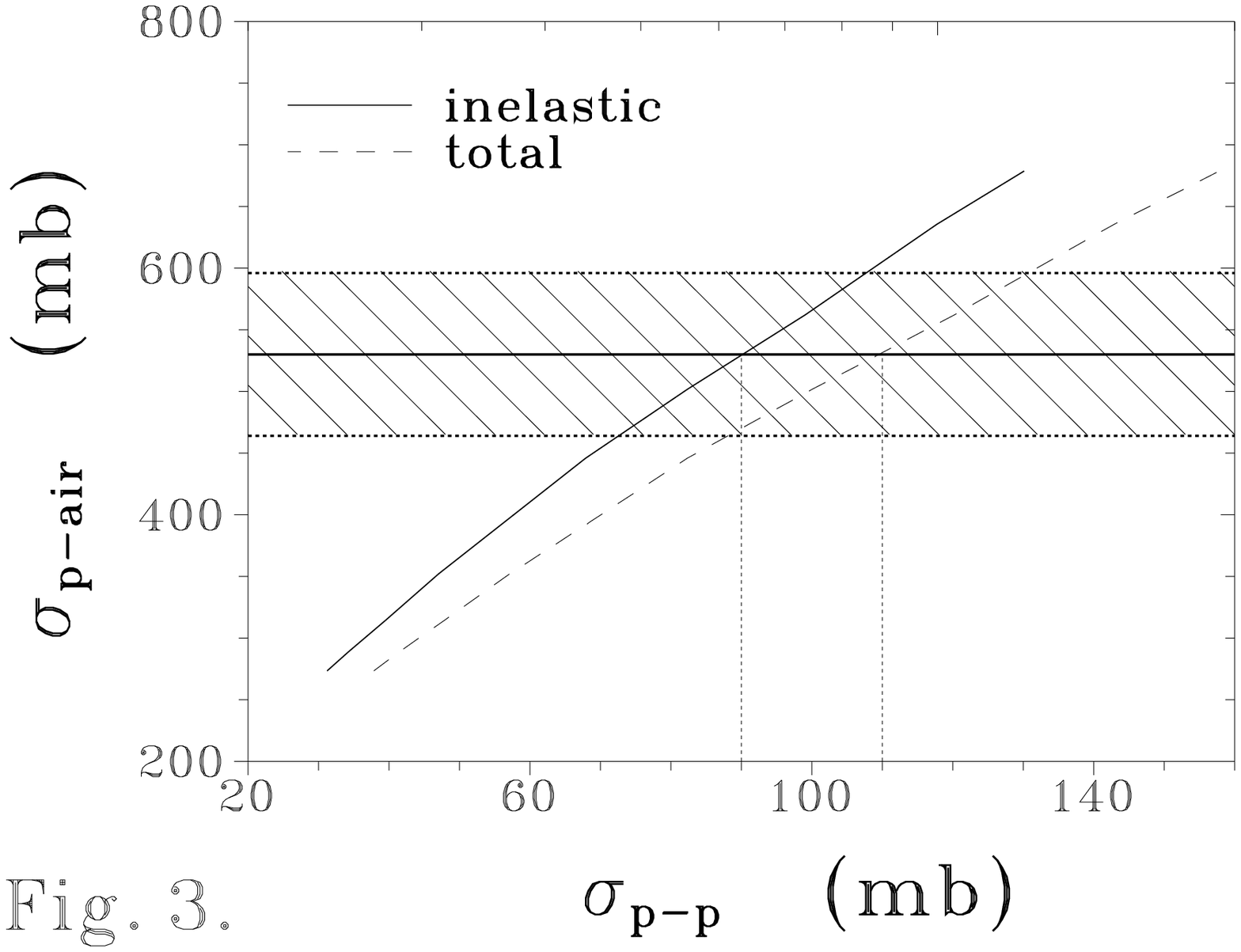,width=8.5cm}
\hspace{-1.cm}
%
\psfig{file=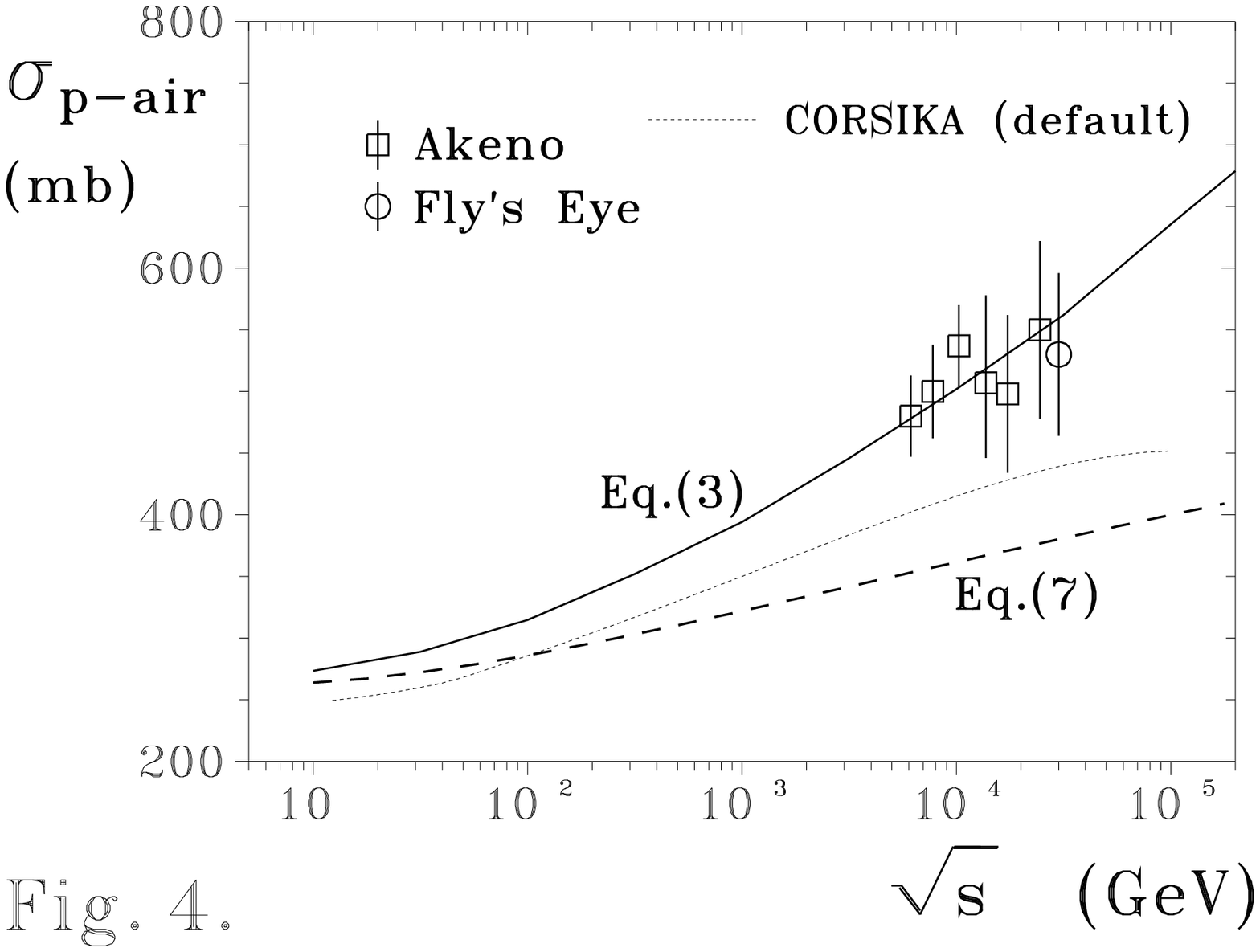,width=8.5cm}}
\vspace{.8cm}


\vspace{.3cm}
Fig.1.
{Nitrogen nucleus profile ( phase shift: $\chi_{p-{\rm N}}$)
obtained using the
exact Glauber formula with geometrical scaling (solid lines)
and factorization hypothesis
(simplified Glauber) (dashed lines) for
different interaction energy (per nucleon--nucleon collision)}
\vspace{.3cm}

Fig.2.
Cross section of collisions
of different nuclei with the ``air nucleus'' calculated
using exact Glauber formula with geometrical scaling
(solid lines) and simplified formulas (dashed lines)
as a function of interaction energy.
\vspace{.3cm}

{Fig.3.
Relationship between inelastic proton--air cross section
and the value of proton--proton cross section (inelastic - solid curve
and total - dashed one)
calculated using exact Glauber formula with geometrical scaling.
Solid horizontal line represent the value measured in Fly's Eye [2]
experiment (dashed area shows 1$\sigma$ bounds).}
\vspace{.3cm}

{Fig.4. Inelastic proton--air cross sections
calculated using exact Glauber formula with geometrical scaling
as a function of interaction energy (per nucleus).
Experimental points are from Akeno and Fly's Eye experiments
(squares and the circle, respectively). Results of
calculations with the ``simplified Glauber'' formula with the
same proton--proton cross section energy dependence are given by the
long dashed line. The default CORSIKA \cite{corsika} proton--air
cross section is shown for a comparison by the short dashed line.}


\begin{thebibliography}{9}

\bibitem{akeno}
M. Honda {\it et al.}, Phys. Rev. Lett. {\bf 70}, 525 (1993).
\bibitem{FE}
R. M. Baltrusatis {\it et al.}, Phys. Rev. Lett. {bf 52}, 1380 (1984).
\bibitem{glauber}
R. Glauber, in {\it Lectures in Theoretical Physics}, edited by A. O. Barut and
W. E. Brittin (Interscience, New York, 1956).
\bibitem{amaldi}U. Amaldi and K. R. Schubert, Nucl. Phys. B {\bf 166}, 301 (1980).
\bibitem{buddd}A. J. Buras and J. Dias de Deus, Nucl. Phys. B {\bf 71}, 481 (1974).

B. M. Bobchenko {\it et al.}, Sov. J. Nucl. Phys. {\bf 30}, 805 (1979);
S. Fredriksson, G. Eilam, G. Berlad and L. Bergstr\"{o}m,
Phys. Rep. {\bf 144}, 187 (1987).

\bibitem{xsecprep}
T. Wibig and D. Sobczy\'{n}ska, in preparation.

\bibitem{corsika}
D. Heck, J. Knapp, J. N. Capdevielle, G. Schatz and T. Thouw,
Forschungszentrum Karlsruhe Report No. FZKA 6019, Karlsruhe, (1998).



\end{thebibliography}
\end{document}